\documentclass[11pt]{article}
\usepackage{amssymb,latexsym,amsmath,amsbsy,amsthm}
\usepackage[dvips]{graphicx}
\usepackage{xcolor}
\usepackage{cite}
\usepackage{stmaryrd}  % for \llbracket and \rrbracket
\def\nn{\nonumber}
\def\deg{\mathop{\rm deg}\nolimits}

\def\qdots{\mathinner{\mkern1mu\raise1pt\vbox{\kern7pt\hbox{.}}\mkern2mu \raise4pt\hbox{.}\mkern2mu\raise7pt\hbox{.}\mkern1mu}}
\def\Z{{\mathbb Z}}

\def\gl{\mathfrak{gl}}
\def\ssl{\mathfrak{sl}}

\def\g{\mathfrak{g}}

\def\so{\mathfrak{so}}

\def\osp{\mathfrak{osp}}
\def\pso{\mathfrak{pso}}
\def\lb{\llbracket}
\def\rb{\rrbracket}

\DeclareMathOperator{\Str}{Str}
\DeclareMathOperator{\tr}{tr}
\DeclareMathOperator{\End}{End}
\renewcommand{\theequation}{\arabic{section}.{\arabic{equation}}}

\begin{document}
\begin{center}
{\Large \bf
Orthosymplectic $\Z_2\times\Z_2$-graded Lie superalgebras\\[3mm] and parastatistics} \\[5mm]
{\bf N.I.~Stoilova}\footnote{E-mail: stoilova@inrne.bas.bg}\\[1mm] 
Institute for Nuclear Research and Nuclear Energy, Bulgarian Academy of Sciencies,\\ 
Boul.\ Tsarigradsko Chaussee 72, 1784 Sofia, Bulgaria\\[2mm] 
{\bf J.\ Van der Jeugt}\footnote{E-mail: Joris.VanderJeugt@UGent.be}\\[1mm]
Department of Applied Mathematics, Computer Science and Statistics, Ghent University,\\
Krijgslaan 281-S9, B-9000 Gent, Belgium.
\end{center}

%\addtolength{\baselineskip}{2mm}
%\addtolength{\abovedisplayskip}{1mm}
%\addtolength{\belowdisplayskip}{1mm}
%\addtolength{\parskip}{2mm}
\vskip 2 cm

\begin{abstract}
\noindent 
A ${\mathbb Z}_2\times{\mathbb Z}_2$-graded Lie superalgebra $\mathfrak g$ is a ${\mathbb Z}_2\times{\mathbb Z}_2$-graded algebra with a bracket $\llbracket \cdot , \cdot \rrbracket$ that satisfies certain graded versions of the symmetry and Jacobi identity. 
In particular, despite the common terminology, $\mathfrak g$ is not a Lie superalgebra.
We construct the most general orthosymplectic ${\mathbb Z}_2\times{\mathbb Z}_2$-graded Lie superalgebra $\osp(2m_1+1,2m_2|2n_1,2n_2)$ in terms of defining matrices.
A special case of this algebra appeared already in work of Tolstoy in 2014.
Our construction is based on the notion of graded supertranspose for a $\Z_2\times\Z_2$-graded matrix.
Since the orthosymplectic Lie superalgebra $\osp(2m+1|2n)$ is closely related to the definition of parabosons, parafermions and mixed parastatistics, we investigate here the new parastatistics relations following from $\osp(2m_1+1,2m_2|2n_1,2n_2)$.
Some special cases are of particular interest, even when one is dealing with parabosons only. 
\end{abstract}

\vskip 10mm
\noindent $\Z_2\times\Z_2$-graded Lie superalgebras 


%%%%%%%%%%%%%%%%%%%%%%%%%%%%%%%%%%%%%%%%%%%%%%%%%%%%%%%%%%%%%%%%%%%%%%%%%%%%%%%
%%%%% section
%%%%%%%%%%%%%%%%%%%%%%%%%%%%%%%%%%%%%%%%%%%%%%%%%%%%%%%%%%%%%%%%%%%%%%%%%%%%%%%
\setcounter{equation}{0}
\section{Introduction} \label{sec:A}%

The orthosymplectic Lie superalgebra $\osp(2m+1|2n)$ is known to be closely related to parastatistics.
In fact, specific generators of the algebra describe a system of $m$ parafermions and $n$ parabosons that satisfy the so-called relative parafermion relations~\cite{Palev1982}.
Already from the early days of parastatistics~\cite{Green}, it was known that such systems of $m$ parafermions and $n$ parabosons can also satisfy a different type of relative relations, the so-called relative paraboson relations~\cite{GM}.
The algebra generated by this second system is no longer a Lie superalgebra, but a $\Z_2\times\Z_2$-graded Lie superalgebra, as was discovered by Tolstoy~\cite{Tolstoy2014}.
Such $\Z_2\times\Z_2$-graded Lie superalgebras belong to the more general class of colour superalgebras~\cite{Rit1,Rit2,Scheunert1979}.
The term $\Z_2\times\Z_2$-graded Lie superalgebra is somewhat misleading, since in general this is not a Lie superalgebra.
But since this term is now commonly used, we shall also adopt it.

Colour algebras and colour superalgebras were introduced by Rittenberg and Wyler~\cite{Rit1,Rit2}.
Such algebras are graded by some abelian grading group $\Gamma$, and the simplest case not coinciding with a Lie superalgebra is for $\Gamma=\Z_2\times\Z_2$.
For an algebra graded by $\Z_2\times\Z_2$, there are already two distinct choices for the Lie bracket: one corresponding to $\Z_2\times\Z_2$-graded Lie algebras and one corresponding to $\Z_2\times\Z_2$-graded Lie superalgebras.
Applications of $\Z_2\times\Z_2$-graded Lie superalgebras in physics were rather uncommon~\cite{LR1978,Vasiliev1985,JYW1987}.
Since the recognition of a $\Z_2\times\Z_2$-graded Lie superalgebra underlying the symmetries of L\'evy–Leblond equations~\cite{Aizawa1,Aizawa3}, these $\Z_2\times\Z_2$-graded algebras have experienced a revival in mathematical physics.
They appeared in graded (quantum) mechanics and quantization~\cite{Bruce2,AMD2020,Aizawa4,Aizawa5,Quesne2021,Aizawa6},
in $\Z_2\times\Z_2$-graded two-dimensional models~\cite{Bruce1,Toppan1,Bruce3},
in $\Z_2\times\Z_2$-graded superspace formulations~\cite{Poncin,Doi,Aizawa7,Aizawa8}
and in particular in parastatistics~\cite{Tolstoy2014,SV2018} and in the description and application of other types of parabosons and parafermions~\cite{Toppan2,Toppan3}.

In this paper, we will focus on orthosymplectic $\Z_2\times\Z_2$-graded Lie superalgebras, and their relation to parastatistics.
As mentioned, the first such interesting case was discovered by Tolstoy~\cite{Tolstoy2014}.
In his notation, the systems consisting of $m$ parafermions and $n$ parabosons satisfying the relative paraboson relations is described by the $\Z_2\times\Z_2$-graded Lie superalgebra $\osp(1,2m|2n,0)$. 
Parastatistics Fock spaces for this system were studied in~\cite{SV2018}, where this particular algebra was denoted by $\pso(2m+1|2n)$.
In Tolstoy's paper, a more general situation is considered: the $\Z_2\times\Z_2$-graded Lie superalgebra $\osp(2m_1+1,2m_2|2n,0)$ is defined, and related to a system with two types of parafermions and one type of parabosons.
The purpose of the current paper is to describe the most general orthosymplectic $\Z_2\times\Z_2$-graded Lie superalgebra, denoted by $\osp(2m_1+1,2m_2|2n_1,2n_2)$, and to consider the corresponding parastatistics system.

In Section~\ref{sec:B} we first recall some general definitions for $\Z_2\times\Z_2$-graded Lie superalgebras, and give the simplest example $\gl(m_1,m_2|n_1,n_2)$ (and closely related $\ssl(m_1,m_2|n_1,n_2)$).
In order to define $\osp(2m_1+1,2m_2|2n_1,2n_2)$ in matrix form, it is useful to remind how the ordinary Lie superalgebra $\osp(2m+1|2n)$ is imbedded in the Lie superalgebra $\gl(2m+1|2n)$, which is presented in Section~\ref{sec:C}.
Then we explain how $\osp(2m_1+1,2m_2|2n_1,2n_2)$ is imbedded in $\gl(2m_1+1,2m_2|2n_1,2n_2)$ (as $\Z_2\times\Z_2$-graded Lie superalgebras). 
Our matrix form looks fairly involved, but is most appropriate for the identification of paraboson and parafermion generators.
Our matrices for $\osp(2m_1+1,2m_2|2n_1,0)$ are also different from the ones appearing in~\cite{Tolstoy2014}, but essentially this is only up to a reordering of the row and column indices.
The main point is that we present the most general orthosymplectic $\Z_2\times\Z_2$-graded Lie superalgebra (not only for $n_2=0$ as in\cite{Tolstoy2014}).
Furthermore, by introducing a graded supertranspose we can characterize $\osp(2m_1+1,2m_2|2n_1,2n_2)$ in a transparent form.

The basic classical Lie superalgebras consist (apart from exceptional ones) of four series, of type $A$, $B$, $C$ and $D$~\cite{Kac1977}.
The algebras of type $A$ are $\ssl(m|n)$, and those of type $B$, $C$ and $D$ are orthosymplectic, namely $\osp(2m+1|2n)$, $\osp(2|2n)$ and $\osp(2m|2n)$. 
Since it is fairly trivial to recognize a $\Z_2\times\Z_2$-graded subalgebra $\osp(2m_1,2m_2|2n_1,2n_2)$ of the $\Z_2\times\Z_2$-graded Lie superalgebra $\osp(2m_1+1,2m_2|2n_1,2n_2)$, we have now $\Z_2\times\Z_2$-graded versions for all of these series.

In Section~\ref{sec:D}, we present a number of applications in terms of parastatistics operators.
Just as the parafermion operators and paraboson operators are identified with root vectors of $\so(2m+1)$ or $\osp(1|2n)$ (in particular, the root vectors corresponding to the short roots of the Lie algebra or Lie superalgebra), we identify here the corresponding generators of the $\Z_2\times\Z_2$-graded Lie superalgebra $\osp(2m_1+1,2m_2|2n_1,2n_2)$.
This gives rise to a system of two families of parabosons and two families of parafermions, with mutual triple relations.
Note that these four families appear because of the $\Z_2\times\Z_2$ grading.
For the most general case, the triple relations determining the parastatistics is given in the Appendix.
A special case, namely for $\osp(1,0|2n_1,2n_2)$, is given as an interesting example in Section~\ref{sec:D}.
In the same section, we also briefly present another type of parastatistics relations, namely those related to the 
$\Z_2\times\Z_2$-graded Lie superalgebra $\ssl(1,0|n_1,n_2)$.
Some final remarks are summarized in Section~\ref{sec:E}.

%%%%%%%%%%%%%%%%%%%%%%%%%%%%%%%%%%%%%%%%%%%%%%%%%%%%%%%%%%%%%%%%%%%%%%%%%%%%%%%
%%%%% section
%%%%%%%%%%%%%%%%%%%%%%%%%%%%%%%%%%%%%%%%%%%%%%%%%%%%%%%%%%%%%%%%%%%%%%%%%%%%%%%
\section{$\Z_2\times\Z_2$-graded Lie superalgebras and $\gl(m_1,m_2|n_1,n_2)$}
\setcounter{equation}{0} \label{sec:B}

The definition of $\Z_2\times\Z_2$-graded Lie algebras and $\Z_2\times\Z_2$-graded Lie superalgebras ($\Z_2^2$-GLSA) 
as well as examples of such algebras goes back to papers of Rittenberg and Wyler~\cite{Rit1,Rit2}.

The $\Z_2^2$-GLSA $\g$, as a linear space, is a direct sum of four subspaces:
\begin{equation}
\g=\bigoplus_{\boldsymbol{a}} \g_{\boldsymbol{a}} =
\g_{(0,0)} \oplus \g_{(0,1)} \oplus \g_{(1,0)} \oplus \g_{(1,1)} 
\label{ZZgrading}
\end{equation}
where $\boldsymbol{a}=(a_1,a_2)$ is an element of $\Z_2\times\Z_2$.
Elements of $\g_{\boldsymbol{a}}$ are denoted by $x_{\boldsymbol{a}}, y_{\boldsymbol{a}},\ldots$,
and $\boldsymbol{a}$ is called the degree, $\deg x_{\boldsymbol{a}}$, of $x_{\boldsymbol{a}}$.
Such elements are called homogeneous elements.
The $\Z_2\times\Z_2$-graded Lie superalgebra $\g$ admits a bilinear operation  $\lb\cdot,\cdot\rb$ which
satisfies the grading, symmetry and Jacobi identities:
\begin{align}
& \lb x_{\boldsymbol{a}}, y_{\boldsymbol{b}} \rb \in \g_{\boldsymbol{a}+\boldsymbol{b}}, \label{grading}\\
& \lb x_{\boldsymbol{a}}, y_{\boldsymbol{b}} \rb = -(-1)^{\boldsymbol{a}\cdot\boldsymbol{b}} 
\lb y_{\boldsymbol{b}}, x_{\boldsymbol{a}} \rb, \label{symmetry}\\
& \lb x_{\boldsymbol{a}}, \lb y_{\boldsymbol{b}}, z_{\boldsymbol{c}}\rb \rb =
\lb \lb x_{\boldsymbol{a}}, y_{\boldsymbol{b}}\rb , z_{\boldsymbol{c}} \rb +
(-1)^{\boldsymbol{a}\cdot\boldsymbol{b}} \lb y_{\boldsymbol{b}}, \lb x_{\boldsymbol{a}}, z_{\boldsymbol{c}}\rb \rb,
\label{jacobi}
\end{align} 
where
\begin{equation}
\boldsymbol{a}+\boldsymbol{b}=(a_1+b_1,a_2+b_2)\in \Z_2\times\Z_2, \qquad
\boldsymbol{a}\cdot\boldsymbol{b} = a_1b_1+a_2b_2.
\label{sign}
\end{equation}
Observe that another choice for $\boldsymbol{a}\cdot\boldsymbol{b}$, namely $\boldsymbol{a}\cdot\boldsymbol{b} = a_1b_2-a_2b_1$, 
would yield the definition of a $\Z_2\times\Z_2$-graded Lie algebra~\cite{Rit2,SV2023}.

It is important to understand that in general a $\Z_2\times\Z_2$-graded Lie superalgebra is not a Lie superalgebra, since the bracket properties are different.

Let $\g$ be an associative algebra with a product denoted by $x\cdot y$, and suppose $\g$ has a $\Z_2\times\Z_2$-grading of the form~\eqref{ZZgrading} that is compatible with the product, i.e.\ $x_{\boldsymbol{a}} \cdot y_{\boldsymbol{b}} \in \g_{\boldsymbol{a}+\boldsymbol{b}}$, then the following bracket turns $\g$ into a $\Z_2^2$-GLSA:
\begin{equation}
\lb x_{\boldsymbol{a}} , y_{\boldsymbol{b}}\rb = x_{\boldsymbol{a}} \cdot y_{\boldsymbol{b}}- 
(-1)^{\boldsymbol{a}\cdot\boldsymbol{b}} y_{\boldsymbol{b}} \cdot x_{\boldsymbol{a}}\ ,
\label{ZZbracket}
\end{equation}
i.e.\ for a bracket derived from an associative product the Jacobi identity~\eqref{jacobi} is automatically satisfied.

The generic example is $\gl(m_1,m_2|n_1,n_2)$~\cite{Rit1,Rit2,Tolstoy2014,Isaac2020}. 
Let $V$ be a $\Z_2\times \Z_2$-graded linear space, 
{$V=V_{(0,0)} \oplus V_{(1,1)} \oplus V_{(1,0)} \oplus V_{(0,1)}$},
with subspaces of dimension $m_1,m_2,n_1$ and $n_2$ respectively.
{$\End(V)$ is then a $\Z_2\times \Z_2$-graded associative algebra, 
and by the previous property it is turned into a $\Z_2^2$-GLSA by the bracket $\lb\cdot,\cdot\rb$ of~\eqref{ZZbracket}. 
This algebra is usually denoted by $\gl(m_1,m_2|n_1,n_2)$. 
In matrix form, the elements are written as:
\begin{equation}
A = \begin{array}{c c}
    {\begin{array} {@{} c c  cc @{}}  m_1\ & \ m_2\ & \ n_1\ & \ \ n_2 \ \end{array} } & {} \\  %line1
    \left(\begin{array}{cccc} 
a_{(0,0)} & a_{(1,1)} & a_{(1,0)} & a_{(0,1)} \\[1mm] 
b_{(1,1)} & b_{(0,0)} & b_{(0,1)} & b_{(1,0)} \\[1mm] 
c_{(1,0)} & c_{(0,1)} & c_{(0,0)} & c_{(1,1)} \\[1mm] 
d_{(0,1)} & d_{(1,0)} & d_{(1,1)} & d_{(0,0)} 
    \end{array}\right)
 & \hspace{-2mm} %\hspace{-1em}
		{\begin{array}{l}
     m_1 \\[1mm]  m_2 \\[1mm]	n_1 \\[1mm] n_2
    \end{array} } \\ %line2 
    %\mbox{} % Blank line to match column names so as to align the = vertically
  \end{array} %\\[-12pt]  . % Correction for blank line
\label{ZZsl}
\end{equation}
The indices of the matrix blocks refer to the $\Z_2\times\Z_2$-grading, and the size of the blocks is indicated in the lines above and to the right of the matrix.

The matrices of the Lie algebra $\gl(m_1+m_2+n_1+n_2)$, of the Lie superalgebra $\gl(m_1+m_2|n_1+n_2)$ and of the $\Z_2\times\Z_2$-graded Lie superalgebra $\gl(m_1,m_2|n_1,n_2)$ are all the same, but of course the bracket is different in all of these cases.

One can check that $\Str \lb A,B \rb =0$, where $\Str(A)=\tr(a_{(0,0)})+\tr(b_{(0,0)})-\tr(c_{(0,0)})-\tr(d_{(0,0)})$ is the graded supertrace in terms of the ordinary trace $\tr$.
Hence $\ssl(m_1,m_2|n_1,n_2)$ is defined as the subalgebra of elements $\gl(m_1,m_2|n_1,n_2)$ with graded supertrace equal to~0.

Note that the row and column indices $(i,j)$ of the matrix $A$ can be simultaneously permuted $(i,j\in\{1,2,\ldots,m_1+m_2+n_1+n_2\})$: this would not change the $\Z_2^2$-GLSA itself, but the separation of the blocks of homogeneous elements would not be as simple as in~\eqref{ZZsl}.
For the introduction of an appropriate form of the orthosymplectic $\Z_2^2$-GLSA's, however, it will be necessary to do this.

An important notion to be introduced is that of graded supertranspose, for which we follow a definition similar to that of supertranspose for Lie superalgebras as in~\cite{Scheunert}.
Let $A$ be a homogeneous element of $\ssl(m_1,m_2|n_1,n_2) \subset \End(V)$ of degree $\boldsymbol{a}\in\Z_2\times\Z_2$.
Let $V^*$ be the vector space dual to $V$, inheriting the $\Z_2\times\Z_2$-grading from $V$, 
and denote the natural pairing of $V$ and $V^*$ by $\langle\cdot,\cdot\rangle$.
Then $A^* \in \End(V^*)$ is determined by:
\begin{equation}
\langle A^* y_{\boldsymbol{b}},x\rangle= (-1)^{{\boldsymbol{a}}\cdot{\boldsymbol{b}}} \langle y_{\boldsymbol{b}}, Ax \rangle,
\qquad \forall y_{\boldsymbol{b}}\in V^*_{\boldsymbol{b}}, \forall x\in V.
\label{defT}
\end{equation}
This is extended by linearity to all elements of $\ssl(m_1,m_2|n_1,n_2)$.
In matrix form, this yields the {$\Z_2\times\Z_2$-graded supertranspose $A^T$} of $A$:
\begin{equation}
A^T=\left(\begin{array}{cccc} 
a_{(0,0)}^t & b_{(1,1)}^t & -c_{(1,0)}^t & -d_{(0,1)}^t \\ 
a_{(1,1)}^t & b_{(0,0)}^t & c_{(0,1)}^t & d_{(1,0)}^t \\ 
a_{(1,0)}^t & -b_{(0,1)}^t & c_{(0,0)}^t & -d_{(1,1)}^t \\ 
a_{(0,1)}^t & -b_{(1,0)}^t & -c_{(1,1)}^t & d_{(0,0)}^t 
    \end{array}\right),
\label{AT}		
\end{equation}
where $a^t$ denotes the ordinary matrix transpose.
One can check (case by case, according to the $\Z_2\times\Z_2$-grading) that the graded supertranspose of matrices satisfies
\begin{equation}
(AB)^T = (-1)^{{\boldsymbol{a}}\cdot{\boldsymbol{b}}} B^T A^T,
\label{ABT}
\end{equation}
where the sign is determined by~\eqref{sign}.

%%%%%%%%%%%%%%%%%%%%%%%%%%%%%%%%%%%%%%%%%%%%%%%%%%%%%%%%%%%%%%%%%%%%%%%%%%%%%%%%%%%%%%%%
\setcounter{equation}{0}
\section{The $\Z_2\times\Z_2$-graded Lie superalgebra $\osp(2m_1+1,2m_2|2n_1,2n_2)$} 
\label{sec:C}%

Before we turn to the matrix form of $\osp(2m_1+1,2m_2|2n_1,2n_2)$, it is helpful to recall the standard matrix form of the Lie superalgebra $\osp(2m+1|2n)$, as a subalgebra of the Lie superalgebra $\ssl(2m+1|2n)$~\cite{Kac1977,Scheunert,SV2015}.

In general, the Lie superalgebra $\gl(m|n)$ consists of $\left( \begin{array}{cc} a&b\\ c&d \end{array}\right)$ with $a$ an $m\times m$-matrix, $d$ an $n\times n$-matrix, and $b$ and $c$ rectangular matrices of appropriate size. Matrices of the form $\left( \begin{array}{cc} a&0\\ 0&d \end{array}\right)$ are even elements and matrices of the form $\left( \begin{array}{cc} 0&b\\ c&0 \end{array}\right)$ are odd elements of the Lie superalgebra. The Lie superalgebra $\ssl(m|n)$ consists of such matrices with supertrace zero, i.e.\ with $\tr a- \tr d =0$.
For Lie superalgebras, there is also the notion of supertranspose~\cite{Scheunert}. 
For an element of the form $A=\left( \begin{array}{cc} a&b\\ c&d \end{array}\right)$, 
this supertranspose is $A^{ST}=\left( \begin{array}{cc} a^t&-c^t\\ b^t&d^t \end{array}\right)$.

The Lie superalgebra $\osp(2m+1|2n)$ is a subalgebra of $\ssl(2m+1|2n)$ consisting of the set of matrices of the following block form:
\begin{equation}
A = 
\begin{array}{c c}
    \begin{array} {@{} c c c cc @{}} \ \ m\ \ & \ \ m\ \ & \ \ 1 \ \ &\ \ n\ \ & \ \ n \ \ \end{array} & {} \\[1mm]  %line1
\left(\begin{array}{ccccc} 
a^{[1,1]} & a^{[1,2]} & a^{[1,3]} & b^{[1,1]} & b^{[1,2]} \\ 
a^{[2,1]} & a^{[2,2]} & a^{[2,3]} & b^{[2,1]} & b^{[2,2]} \\ 
a^{[3,1]} & a^{[3,2]} & a^{[3,3]} & b^{[3,1]} & b^{[3,2]} \\ 
c^{[1,1]} & c^{[1,2]} & c^{[1,3]} & d^{[1,1]} & d^{[1,2]} \\ 
c^{[2,1]} & c^{[2,2]} & c^{[2,3]} & d^{[2,1]} & d^{[2,2]} 
\end{array}\right)
 & \hspace{-1em}
		\begin{array}{l}
     m \\  m \\ 1 \\	n \\ n
    \end{array} \\ %line2 
    \mbox{} % Blank line to match column names so as to align the = vertically
  \end{array} \\[-12pt]   % Correction for blank line
\label{ospmn}
\end{equation}	
such that
\begin{equation}
A^{ST} J + JA=0
\label{AJ1}
\end{equation}
where
\begin{equation}
J =
\left(\begin{array}{ccccc} 
0 & I_m & 0 & 0 & 0 \\ 
I_m & 0 & 0 & 0 & 0 \\ 
0 & 0 & 1 & 0 & 0 \\ 
0 & 0 & 0 & 0 & I_n \\ 
0 & 0 & 0 & -I_n & 0 
\end{array}\right).
\label{J1}
\end{equation}	
Herein, $I_k$ denotes the identity matrix of size $k\times k$ and $0$ stands for a zero block. 
Concretely, this means that the block matrices in~\eqref{ospmn} satisfy
\begin{align}
& a^{[2,2]}=-{a^{[1,1]}}^t,\quad a^{[3,1]}=-{a^{[2,3]}}^t, \quad a^{[3,2]}=-{a^{[1,3]}}^t, \quad a^{[3,3]}=0, \quad d^{[2,2]}=-{d^{[1,1]}}^t, 
\nonumber\\
& a^{[1,2]}\hbox{ and }a^{[2,1]}\hbox{ skew symmetric,} \quad d^{[1,2]}\hbox{ and }d^{[2,1]}\hbox{ symmetric,} \nonumber \\
& c^{[1,1]}={b^{[2,2]}}^t, \quad c^{[1,2]}={b^{[1,2]}}^t, \quad c^{[1,3]}={b^{[3,2]}}^t, \label{abcd1} \\
&c^{[2,1]}=-{b^{[2,1]}}^t, \quad c^{[2,2]}=-{b^{[1,1]}}^t, \quad c^{[2,3]}=-{b^{[3,1]}}^t.\nonumber
\end{align}

Let us now define $\osp(2m_1+1,2m_2|2n_1,2n_2)$ as a subalgebra of $\ssl(2m_1+1,2m_2|2n_1,2n_2)$.
For the matrix form of the $\Z_2^2$-GLSA $\osp(2m_1+1,2m_2|2n_1,2n_2)$, it will be necessary to indicate the $\Z_2\times\Z_2$-grading of the matrix blocks by an index, and as usual the size of the blocks is determined in an obvious way by the borders.
Then our definition is as follows.
The $\Z_2\times\Z_2$-graded Lie superalgebra $\osp(2m_1+1,2m_2|2n_1,2n_2)$ consists of the set of matrices of the following block form:
\begin{equation}
A = 
\begin{array}{c c}
    \begin{array} {@{} c c c c c c c c c @{}} 
		\  m_1\ \ &  \ m_2 \ \ & \  m_1\ \ & \  m_2 \ \ &\  1\ \ &\  n_1\ \ & \  n_2 \ \ & \  n_1\ \ &  \ n_2 \ \ 
		\end{array} & {} \\[1mm]  %line1
\left(\begin{array}{ccccccccc} 
a^{[1,1]}_{(0,0)} & a^{[1,2]}_{(1,1)} & a^{[1,3]}_{(0,0)} & a^{[1,4]}_{(1,1)} & a^{[1,5]}_{(0,0)} 
   & b^{[1,1]}_{(1,0)} & b^{[1,2]}_{(0,1)} & b^{[1,3]}_{(1,0)} & b^{[1,4]}_{(0,1)} \\[2mm] 
a^{[2,1]}_{(1,1)} & a^{[2,2]}_{(0,0)} & a^{[2,3]}_{(1,1)} & a^{[2,4]}_{(0,0)} & a^{[2,5]}_{(1,1)} 
   & b^{[2,1]}_{(0,1)} & b^{[2,2]}_{(1,0)} & b^{[2,3]}_{(0,1)} & b^{[2,4]}_{(1,0)} \\[2mm]  
a^{[3,1]}_{(0,0)} & a^{[3,2]}_{(1,1)} & a^{[3,3]}_{(0,0)} & a^{[3,4]}_{(1,1)} & a^{[3,5]}_{(0,0)} 
   & b^{[3,1]}_{(1,0)} & b^{[3,2]}_{(0,1)} & b^{[3,3]}_{(1,0)} & b^{[3,4]}_{(0,1)} \\[2mm] 
a^{[4,1]}_{(1,1)} & a^{[4,2]}_{(0,0)} & a^{[4,3]}_{(1,1)} & a^{[4,4]}_{(0,0)} & a^{[4,5]}_{(1,1)} 
   & b^{[4,1]}_{(0,1)} & b^{[4,2]}_{(1,0)} & b^{[4,3]}_{(0,1)} & b^{[4,4]}_{(1,0)} \\[2mm] 
a^{[5,1]}_{(0,0)} & a^{[5,2]}_{(1,1)} & a^{[5,3]}_{(0,0)} & a^{[5,4]}_{(1,1)} & a^{[5,5]}_{(0,0)} 
   & b^{[5,1]}_{(1,0)} & b^{[5,2]}_{(0,1)} & b^{[5,3]}_{(1,0)} & b^{[5,4]}_{(0,1)} \\[2mm] 
c^{[1,1]}_{(1,0)} & c^{[1,2]}_{(0,1)} & c^{[1,3]}_{(1,0)} & c^{[1,4]}_{(0,1)} & c^{[1,5]}_{(1,0)} 
   & d^{[1,1]}_{(0,0)} & d^{[1,2]}_{(1,1)} & d^{[1,3]}_{(0,0)} & d^{[1,4]}_{(1,1)} \\[2mm] 
c^{[2,1]}_{(0,1)} & c^{[2,2]}_{(1,0)} & c^{[2,3]}_{(0,1)} & c^{[2,4]}_{(1,0)} & c^{[2,5]}_{(0,1)} 
   & d^{[2,1]}_{(1,1)} & d^{[2,2]}_{(0,0)} & d^{[2,3]}_{(1,1)} & d^{[2,4]}_{(0,0)} \\[2mm] 
c^{[3,1]}_{(1,0)} & c^{[3,2]}_{(0,1)} & c^{[3,3]}_{(1,0)} & c^{[3,4]}_{(0,1)} & c^{[3,5]}_{(1,0)} 
   & d^{[3,1]}_{(0,0)} & d^{[3,2]}_{(1,1)} & d^{[3,3]}_{(0,0)} & d^{[3,4]}_{(1,1)} \\[2mm] 
c^{[4,1]}_{(0,1)} & c^{[4,2]}_{(1,0)} & c^{[4,3]}_{(0,1)} & c^{[4,4]}_{(1,0)} & c^{[4,5]}_{(0,1)} 
   & d^{[4,1]}_{(1,1)} & d^{[4,2]}_{(0,0)} & d^{[4,3]}_{(1,1)} & d^{[4,4]}_{(0,0)} 
\end{array}\right)
 & \hspace{-1em}
		\begin{array}{l}
     m_1 \\[3mm]  m_2 \\[3mm] m_1 \\[3mm] m_2 \\[3mm] 1 \\[3mm]	n_1 \\[3mm] n_2 \\[3mm] n_1 \\[3mm] n_2
    \end{array} \\ %line2 
    \mbox{} % Blank line to match column names so as to align the = vertically
  \end{array} \\[-12pt]   % Correction for blank line
\label{ospmn2}
\end{equation}	
such that
\begin{equation}
A^T J + JA=0
\label{AJ2}
\end{equation}
where
\begin{equation}
J =
\left(\begin{array}{ccccc} 
0 & I_{m_1+m_2} & 0 & 0 & 0 \\ 
I_{m_1+m_2} & 0 & 0 & 0 & 0 \\ 
0 & 0 & 1 & 0 & 0 \\ 
0 & 0 & 0 & 0 & I_{n_1+n_2} \\ 
0 & 0 & 0 & -I_{n_1+n_2} & 0 
\end{array}\right).
\label{J2}
\end{equation}	
In~\eqref{AJ2}, $A^T$ is the graded supertranspose of $A$, determined by~\eqref{AT}. 
Note however that the row and column indices of the matrix in~\eqref{ospmn2} as an element of $\ssl(2m_1+1,2m_2|2n_1,2n_2)$ have been appropriately permuted. 
This was done in order to preserve an analogy with the matrices of $\osp(2m+1|2n)$, and in order to have a proper relation with parafermions and parabosons in a later section.

The $\Z_2\times\Z_2$-graded Lie superalgebra $\osp(2m_1+1,2m_2|2n_1,2n_2)$ is well determined by the above definition: this follows from the fact that if such matrices $A$ and $B$ satisfy~\eqref{AJ2}, then their graded bracket $\lb A,B \rb$ also satisfies~\eqref{AJ2}, which is easily deduced from~\eqref{ABT}.

More concretely, $\osp(2m_1+1,2m_2|2n_1,2n_2)$ consists of matrices of the form~\eqref{ospmn2} which satisfy
\begin{align}
& a^{[3,3]}_{(0,0)}=-{a^{[1,1]}_{(0,0)}}^t,\quad a^{[3,4]}_{(1,1)}=-{a^{[2,1]}_{(1,1)}}^t, \quad a^{[4,3]}_{(1,1)}=-{a^{[1,2]}_{(1,1)}}^t, \quad a^{[4,4]}_{(0,0)}=-{a^{[2,2]}_{(0,0)}}^t, \nonumber\\
& a^{[2,3]}_{1,1)}=-{a^{[1,4]}_{(1,1)}}^t,\quad a^{[4,1]}_{(1,1)}=-{a^{[3,2]}_{(1,1)}}^t; \quad a^{[1,3]}_{(0,0)}, a^{[2,4]}_{(0,0)}, a^{[3,1]}_{(0,0)}\hbox{ and } a^{[4,2]}_{(0,0)}\hbox{ skew symmetric}, \nonumber\\
& a^{[5,1]}_{(0,0)}=-{a^{[3,5]}_{(0,0)}}^t, \quad a^{[5,2]}_{(1,1)}=-{a^{[4,5]}_{(1,1)}}^t,\quad a^{[5,3]}_{(0,0)}=-{a^{[1,5]}_{(0,0)}}^t, \quad a^{[5,4]}_{(1,1)}=-{a^{[2,5]}_{(1,1)}}^t,\quad a^{[5,5]}_{(0,0)}=0, \nonumber \\
& d^{[3,3]}_{(0,0)}=-{d^{[1,1]}_{(0,0)}}^t,\quad d^{[3,4]}_{(1,1)}={d^{[2,1]}_{(1,1)}}^t, \quad d^{[4,3]}_{(1,1)}={d^{[1,2]}_{(1,1)}}^t, \quad d^{[4,4]}_{(0,0)}=-{d^{[2,2]}_{(0,0)}}^t, \nonumber\\
& d^{[2,3]}_{1,1)}=-{d^{[1,4]}_{(1,1)}}^t,\quad d^{[4,1]}_{(1,1)}=-{d^{[3,2]}_{(1,1)}}^t; \quad d^{[1,3]}_{(0,0)}, d^{[2,4]}_{(0,0)}, d^{[3,1]}_{(0,0)}\hbox{ and } d^{[4,2]}_{(0,0)}\hbox{ symmetric}, \label{abcd2} \\
& c^{[1,1]}_{(1,0)}={b^{[3,3]}_{(1,0)}}^t, \quad c^{[1,2]}_{(0,1)}=-{b^{[4,3]}_{(0,1)}}^t,\quad c^{[1,3]}_{(1,0)}={b^{[1,3]}_{(1,0)}}^t, \quad c^{[1,4]}_{(0,1)}=-{b^{[2,3]}_{(0,1)}}^t,\quad c^{[1,5]}_{(1,0)}={b^{[5,3]}_{(1,0)}}^t , \nonumber\\
& c^{[2,1]}_{(0,1)}={b^{[3,4]}_{(0,1)}}^t, \quad c^{[2,2]}_{(1,0)}=-{b^{[4,4]}_{(1,0)}}^t,\quad c^{[2,3]}_{(0,1)}={b^{[1,4]}_{(0,1)}}^t, \quad c^{[2,4]}_{(1,0)}=-{b^{[2,4]}_{(1,0)}}^t,\quad c^{[2,5]}_{(0,1)}={b^{[5,4]}_{(0,1)}}^t , \nonumber\\
& c^{[3,1]}_{(1,0)}=-{b^{[3,1]}_{(1,0)}}^t, \quad c^{[3,2]}_{(0,1)}={b^{[4,1]}_{(0,1)}}^t,\quad c^{[3,3]}_{(1,0)}=-{b^{[1,1]}_{(1,0)}}^t, \quad c^{[3,4]}_{(0,1)}={b^{[2,1]}_{(0,1)}}^t,\quad c^{[3,5]}_{(1,0)}=-{b^{[5,1]}_{(1,0)}}^t , \nonumber\\
& c^{[4,1]}_{(0,1)}=-{b^{[3,2]}_{(0,1)}}^t, \quad c^{[4,2]}_{(1,0)}={b^{[4,2]}_{(1,0)}}^t,\quad c^{[4,3]}_{(0,1)}=-{b^{[1,2]}_{(0,1)}}^t, \quad c^{[4,4]}_{(1,0)}={b^{[2,2]}_{(1,0)}}^t,\quad c^{[4,5]}_{(0,1)}=-{b^{[5,2]}_{(0,1)}}^t . \nonumber
\end{align}

Although the above matrix conditions look complicated at first sight, they are not difficult to work with.
Let us also mention the following special cases.
\begin{itemize}
\item
When $m_2=n_2=0$, the $\Z_2^2$-GLSA $\osp(2m_1+1,0|2n_1,0)$ just coincides with the ordinary Lie superalgebra $\osp(2m_1+1|2n_1)$ (with appropriate $\Z_2$ grading).
\item
When $m_1=n_2=0$, the $\Z_2^2$-GLSA $\osp(1,2m_2|2n_1,0)$ coincides with the $\Z_2^2$-GLSA denoted by $\pso(2m_2+1|2n_1)$ in~\cite{SV2018}, or (up to a rearrangement of row and column indices) by $\osp(1,2m_2|2n_1,0)$ in~\cite{Tolstoy2014}.
\item
When $n_1=n_2=0$, $\osp(2m_1+1,2m_2|0,0)$ reduces to the Lie algebra $\so(2m_1+2m_2+1)$.
\item
When $m_1=m_2=n_2=0$, $\osp(1,0|2n_1,0)$ reduces to the Lie superalgebra $\osp(1|2n_1)$.
Similarly, when $m_1=m_2=n_2=0$, $\osp(1,0|0,2n_2)$ reduces to the Lie superalgebra $\osp(1|2n_2)$.
Note however that for $m_1=m_2=0$, $\osp(1,0|2n_1,2n_2)$ does not reduce to a Lie algebra or a Lie superalgebra, but remains a $\Z_2^2$-GLSA. This last case is interesting in parastatistics, see Section~\ref{sec:D}.
\end{itemize}

For $n_2=0$ the algebra $\osp(2m_1+1,2m_2|2n_1,0)$ coincides,  up to a rearrangement of row and column indices, with the most general algebra in~\cite{Tolstoy2014}, hence the same notation is used.

Note that the definition of the $\Z_2^2$-GLSA $\osp(2m_1+1,2m_2|2n_1,2n_2)$ with $n_1\ne 0$ and $n_2\ne 0$ is new, and has not been given before.
There is however an equivalent definition (as pointed out to us by a referee), not using the notion of graded supertranspose.
This alternative definition can be found in~\cite{GreenJarvis}.
In order to explain the connection, consider matrices $A$ of size $M\times M$, where $M=2m_1+1+2m_2+2n_1+2n_2$.
For a row or column index $i$, define $d(i)\in\Z_2\times\Z_2$ as follows (compare to the grading of the matrices in~\eqref{ospmn2}):
\[
\begin{array}{l}
d(i)=(0,0),\quad d(m_1+m_2+i)=(0,0)  \quad(i=1,\ldots,m_1);\\
d(2m_1+2m_2+1)=(0,0) ;  \\
d(m_1+i)=(1,1),\quad d(2m_1+m_2+i)=(1,1)   \quad(i=1,\ldots,m_2);\\
d(2m_1+2m_2+1+i)=(1,0),\quad d(2m_1+2m_2+1+n_1+n_2+i)=(1,0)   \quad(i=1,\ldots,n_1);\\
d(2m_1+2m_2+1+n_1+i)=(0,1),\quad d(2m_1+2m_2+1+2n_1+n_2+i)=(0,1)   \quad(i=1,\ldots,n_2).
\end{array}
\]
Next, define a $M\times M$-matrix $u$ by
\[
u_{ij} = (-1)^{d(i)\cdot d(j)},
\]
with $d(i)\cdot d(j)$ as in~\eqref{sign}.
Consider the following $M\times M$ matrices:
\begin{equation}
s_{ij} = \sum_{k=1}^M J_{ik} e_{kj}-u_{ij}\sum_{k=1}^M J_{jk} e_{ki},
\label{sij}
\end{equation}
where $J$ is given by~\eqref{J2}, and as usual $e_{kj}$ is the notation for the $M\times M$ matrix  
with zeros everywhere except a 1 on the intersection of row~$k$ and column~$j$.
The elements $s_{ij}$ ($i,j=1,\ldots,M$) span the $\Z_2\times\Z_2$-graded Lie superalgebras $\osp(2m_1+1,2m_2|2n_1,2n_2)$.
In fact, \eqref{sij} corresponds to equation~(15) of~\cite{GreenJarvis}.
To make the connection with our definition, it is sufficient to check that each matrix $s_{ij}$ satisfies~\eqref{AJ2}, which is a straightforward task.

To complete this section, let us give the $\Z_2\times\Z_2$-graded Lie superalgebras corresponding to the Lie superalgebras of type $C$ and $D$.
In fact, this is quite easy. 
By deleting row $2m_1+2m_2+1$ and column $2m_1+2m_2+1$ in the matrix form~\eqref{ospmn2}, and the corresponding conditions in~\eqref{abcd2}, one obtains the $\Z_2\times\Z_2$-graded Lie superalgebras $\osp(2m_1,2m_2|2n_1,2n_2)$.
These algebras get no further attention in this paper, since no generators can be identified that satisfy the classical triple relations of parabosons or parafermions.

The definition and matrix forms given in this section might give the impression that associating a $\Z_2\times\Z_2$-graded Lie superalgebra to an ordinary Lie superalgebra is a trivial task.
This is definitely not the case.
In general, there is a procedure that works in one way (from a colour Lie (super)algebra to the corresponding Lie (super)algebra), but not in the other way.
This procedure, sometimes referred to as decolouration, is described in~\cite{Campoamor2009}.
It allows -- under certain conditions -- to multiply basis elements of a colour Lie (super)algebra by elements of a certain graded algebra, yielding basis elements of an ordinary Lie (super)algebra of the same dimension.
A general procedure in the opposite way is not known: 
there is a method (also described in~\cite{Campoamor2009}, and already present in~\cite{Rit2} for the $\Z_2\times\Z_2$ case) to construct a $\Z_2\times\Z_2$-graded Lie (super)algebra from an ordinary Lie (super)algebra by taking tensor products,
but the dimension of the constructed algebra is 4 times the dimension of the original algebra.

%%%%%%%%%%%%%%%%%%%%%%%%%%%%%%%%%%%%%%%%%%%%%%%%%%%%%%%%%%%%%%%%%%%%%%%%%%%%%%%%%%%%%%%%%%%%%
\setcounter{equation}{0}
\section{New relations for parabosons and parafermions} 
\label{sec:D}% 

In the original papers, parabosons and parafermions were introduced by means of creation and annihilation operators satisfying certain triple relations with commutators and anti-commutators~\cite{Green,GM}.
A set of $2m$ parafermion operators ($m$ creation and $m$ annihilation operators, referred to as a system of $m$ parafermions) forms a generating set for the Lie algebra $\so(2m+1)$~\cite{KT,Ryan,SV2008}.
Later, it was realized that a set of $2n$ paraboson operators form a generating set for the Lie superalgebra $\osp(1|2n)$~\cite{Ganchev1980}.
A mixed system of $2m$ parafermion operators and $2n$ paraboson operators satisfying relative parafermion relations forms a generating set for the Lie superalgebra $\osp(2m+1|2n)$~\cite{Palev1982}.
And, discovered more recently,
a mixed system of $2m$ parafermion operators and $2n$ paraboson operators satisfying relative paraboson relations form a generating set for the $\Z_2\times\Z_2$-graded Lie superalgebra $\osp(1,2m|2n,0)\cong\pso(2m+1|2n)$~\cite{Tolstoy2014,SV2018}.

The main consideration in~\cite{Tolstoy2014} is the study of a mixed system of two families of parafermions ($m_1$ parafermions of one type and $m_2$ parafermions of second type) with $n$ parabosons, for which Tolstoy has introduced this $\Z_2\times\Z_2$-graded Lie superalgebra $\osp(2m_1+1,2m_2|2n,0)$. 
The contribution of the present paper allows the study of a mixed system consisting of two families of parafermions ($m_1$ parafermions of one type and $m_2$ parafermions of second type) with two families of parabosons ($n_1$ parabosons of one type and $n_2$ parabosons of second type), in such a way that all sorts of relative relations appear.
The actual description of these triple relations is given in the Appendix of this paper.
Here, it is worth emphasizing a particular interesting case.

Consider the $\Z_2\times\Z_2$-graded Lie superalgebra $\g=\osp(1,0|2n_1,2n_2)$. This has a nontrivial $\Z_2\times\Z_2$ grading, i.e.\ it does not collapse to a Lie superalgebra.
As a set of generators for $\osp(1,0|2n_1,2n_2)$, one can take:
\begin{equation}
b_{i}^-= \sqrt{2}(e_{1,i+1}-e_{n_1+n_2+i+1,1}), \quad b_{i}^+=\sqrt{2}(e_{1,n_1+n_2+i+1}+e_{i+1,1}), \qquad i=1,\ldots , n_1+n_2.
\end{equation}
Note that $b^\pm_i \in \g_{(1,0)}$ for $i=1,\ldots,n_1$, and $b^\pm_i \in \g_{(0,1)}$ for $i=n_1+1,\ldots,n_1+n_2$.
These two sets of elements satisfy the common triple relations of parabosons:
\begin{equation}
[\{ b_{ j}^{\xi}, b_{ k}^{\eta}\} , b_{l}^{\epsilon}]= 
(\epsilon -\xi) \delta_{jl} b_{k}^{\eta}  + (\epsilon -\eta) \delta_{kl}b_{j}^{\xi}.
\label{b-rels}
\end{equation}
Herein $\eta, \epsilon, \xi \in\{+,-\}$ (to be interpreted as $+1$ and $-1$), and either $j,k,l\in \{1,2,\ldots,n_1\}$ or else $j,k,l\in \{n_1+1,\ldots,n_1+n_2\}$.
Because of the different $\Z_2\times\Z_2$-grading, we are dealing with two distinct families of parabosons, and this becomes apparent when we consider the mixed relations.
The mixed triple relations between the two families of parabosons have indeed commutators and anti-commutators on different positions, as implied by the $\Z_2\times\Z_2$-graded bracket. These relations read:
\begin{equation}
\{ [b_{ j}^{\xi}, b_{ k}^{\eta}] , b_{l}^{\epsilon}\}= 
-(\epsilon -\xi) \delta_{jl} b_{k}^{\eta}  + (\epsilon -\eta) \delta_{kl}b_{j}^{\xi},
\label{bb-rels}
\end{equation}
where in this case either $j=1,\ldots,n_1$, $k=n_1+1,\ldots,n_1+n_2$, $l=1,\ldots,n_1+n_2$ or else
$j=n_1+1,\ldots,n_1+n_2$, $k=1,\ldots,n_1$, $l=1,\ldots,n_1+n_2$.
As far as we know, this is the first time that such a system of $n_1+n_2$ pairs of parabosons is described.
The two families of parabosons satisfy the traditional triple relations of Green~\cite{Green}.
But in their mutual relations commutators are replaced by anti-commutators and vice versa.

Since we have also used the $\Z_2\times\Z_2$-graded Lie superalgebra $\ssl(m_1,m_2|n_1,n_2)$ in this paper, it is worthwhile to examine another type of parastatistics.
Indeed, A-superstatistics was studied by Palev~\cite{Palev1980A,Palev2003}, and was relevant in certain physical models, in particular lattice models of strongly correlated electrons\cite{Ruck,Zeyher1}.
This A-superstatistics is based on generator relations for the Lie superalgebra $\ssl(1|n)$, and is sometimes referred to as Palev-superstatistics.
As a generalization to the $\Z_2\times\Z_2$-graded case, let us consider the $\Z_2\times\Z_2$-graded Lie superalgebra $\g=\ssl(1,0|n_1,n_2)$,
and define the following generators:
\begin{equation}
a_i^+=e_{i+1,1}, \quad a_i^-=e_{1,i+1} \quad (i=1,2,\ldots, n_1+n_2).
\end{equation}
Since $a_i^\pm \in \g_{(1,0)}$ for $i=1,\ldots,n_1$ and $a_i^\pm \in \g_{(0,1)}$ for $i=n_1+1,\ldots,n_1+n_2$, these generators fall apart in two distinct families.
The traditional relations of A-superstatistics~\cite{Palev2003} hold whenever one stays in one family:
\begin{align}
& \{ a_i^+, a_j^+ \} = \{ a_i^-, a_j^- \} = 0 , \nonumber \\
& [ \{ a_i^+, a_j^- \}, a_k^+ ] = \delta_{jk} a_i^+ - \delta_{ij} a_k^+,\nonumber \\
& [ \{ a_i^+, a_j^- \}, a_k^- ] = -\delta_{ik} a_j^- + \delta_{ij} a_k^- . \label{A1}
\end{align}
Herein, either $i,j,k \in \{1,2,\ldots,n_1\}$ or else $i,j,k \in \{n_1+1,n_1+2,\ldots,n_1+n_2\}$.
The mixed relations between the two families are as follows:
\begin{align}
& [ a_i^+, a_j^+ ] = [ a_i^-, a_j^- ] = 0 , \nonumber \\
& \{ [a_i^+, a_j^- ], a_k^+ \} = \delta_{jk} a_i^+, \nonumber\\
& \{ [ a_i^+, a_j^- ], a_k^- \} = \delta_{ik} a_j^- . \label{A2}
\end{align}
In~\eqref{A2}, either $i\in \{1,2,\ldots,n_1\}$, $j \in \{n_1+1,\ldots,n_1+n_2\}$, $k \in \{1,\ldots,n_1+n_2\}$,
or else $i\in \{n_1+1,\ldots,n_1+n_2\}$, $j \in \{1,2,\ldots,n_1\}$, $k \in \{1,\ldots,n_1+n_2\}$. 

%%%%%%%%%%%%%%%%%%%%%%%%%%%%%%%%%%%%%%%%%%%%%%%%%%%%%%%%%%%%%%%%%%%%%%%%%%%%%%%%%%%%%%%%%%%%%
\setcounter{equation}{0}
\section{Summary and discussion} 
\label{sec:E}% 

The simplest algebras with bracket relations consisting of commutators and anti-commutators that go beyond Lie algebras and Lie superalgebras are 
$\Z_2\times\Z_2$-graded Lie algebras and $\Z_2\times\Z_2$-graded Lie superalgebras~\cite{SV2023}.
Therefore, they can be considered as fundamental structures in mathematical physics.
In this paper, we have introduced the orthosymplectic $\Z_2\times\Z_2$-graded Lie superalgebra $\osp(2m_1+1,2m_2|2n_1,2n_2)$.
The main reason to consider this algebra is its relation to parastatistics, just as the Lie superalgebra $\osp(1|2n)$ and the Lie algebra $\so(2m+1)$ are related to parabosons and parafermions respectively.
By studying a relevant set of generators of $\osp(2m_1+1,2m_2|2n_1,2n_2)$, we were able to identify them with two families of parabosons and two families of parafermions satisfying the most general mutual triple relations.
Already for some special cases, the $\Z_2\times\Z_2$ grading plays an important role.
For example, for $\osp(1,0|2n_1,2n_2)$ there are no parafermions present in the corresponding parastatistics system, but only two families of parabosons.
But even for such systems, the mutual triple relations are new due to the grading property.
It would be interesting to link such parastatistics systems to known physical models, which we hope to study in the future.

In this context, let us already mention that recent investigations have opened again the question of the existence of parastatistics.
In~\cite{Hazzard}, it is shown that nontrivial parastatistics inequivalent to either bosons or fermions can exist in physical systems.
Although their formulation of parastatistics in terms of a four-index tensor is different from the formulation in terms of triple relations, the generalized exclusion principles and free-particle thermodynamics are closely related~\cite{SV2020}.
In any case, these new insights encourage the further study of parastatistics systems.

%%%%%%%%%%%%%%%%%%%%%%%%%%%%%%%%%%%%%%%%%%%%%%%%%%%%%%%%%%%%%%%%%%%%%%%%%%%%%%%%%%%%%%%%%%%%%
\setcounter{equation}{0}
\renewcommand{\theequation}{A.{\arabic{equation}}}
\section*{Appendix} 
\label{sec:F}% 

In this Appendix we list the relations between the parabosons and parafermions in the most general case, following from the $\Z_2^2$-GLSA $\g=\osp(2m_1+1,2m_2|2n_1,2n_2)$.
In terms of the matrix realization~\eqref{ospmn2}, the parafermion creation and annihilation operators are defined as
\begin{align}
&f_i^+=\sqrt{2}(e_{2m_1+2m_2+1,i}-e_{m_1+m_2+i,2m_1+2m_2+1}), \nonumber\\
&f_i^-=\sqrt{2}(e_{i,2m_1+2m_2+1}-e_{2m_1+2m_2+1,m_1+m_2+i}) \quad(i=1,\ldots,m_1+m_2).
\end{align}
Note that $f_i^\pm \in\g_{(0,0)}$ for $i=1,\ldots,m_1$ and $f_i^\pm \in\g_{(1,1)}$ for $i=m_1+1,\ldots,m_1+m_2$, so we are dealing with two families of parafermions.
The paraboson creation and annihilation operators are defined as
\begin{align}
&b_i^+=\sqrt{2}(e_{2m_1+2m_2+1,2m_1+2m_2+1+n_1+n_2+i}+e_{2m_1+2m_2+1+i,2m_1+2m_2+1}), \nonumber\\
&b_i^-=\sqrt{2}(e_{2m_1+2m_2+1,2m_1+2m_2+1+i}-e_{2m_1+2m_2+1+n_1+n_2+i,2m_1+2m_2+1}) \quad(i=1,\ldots,n_1+n_2).
\end{align}
Here $b_i^\pm \in\g_{(1,0)}$ for $i=1,\ldots,n_1$ and $b_i^\pm \in\g_{(0,1)}$ for $i=n_1+1,\ldots,n_1+n_2$, so there are also two families of parabosons.

The parastatistics triple relations for the parafermion creation and annihilation operators $f_i^\pm$ ($i=1,\ldots,m_1+m_2$) do not depend on the family, and coincide with the usual one
\begin{equation}
[[f_{ j}^{\xi}, f_{ k}^{\eta}], f_{l}^{\epsilon}]=
|\epsilon -\eta| \delta_{kl} f_{j}^{\xi} - |\epsilon -\xi| \delta_{jl}f_{k}^{\eta}, 
\label{f-rels}
\end{equation}
where $j,k,l\in\{1,\ldots,m_1+m_2\}$.
Because of the $\Z_2\times\Z_2$-grading and the relevant bracket, the parastatistics triple relations for the paraboson creation and annihilation operators $b_i^\pm$ ($i=1,\ldots,n_1+n_2$) do depend on the family.
If one stays within the same family, one has, as before,
\begin{equation}
[\{ b_{ j}^{\xi}, b_{ k}^{\eta}\} , b_{l}^{\epsilon}]= 
(\epsilon -\xi) \delta_{jl} b_{k}^{\eta}  + (\epsilon -\eta) \delta_{kl}b_{j}^{\xi},
\label{Ab-rels}
\end{equation}
where $j,k,l\in \{1,2,\ldots,n_1\}$ or else $j,k,l\in \{n_1+1,\ldots,n_1+n_2\}$.
For parabosons of the two different families, the mixed relations are, as in the special example of Section~\ref{sec:D}:
\begin{equation}
\{ [b_{ j}^{\xi}, b_{ k}^{\eta}] , b_{l}^{\epsilon}\}= 
-(\epsilon -\xi) \delta_{jl} b_{k}^{\eta}  + (\epsilon -\eta) \delta_{kl}b_{j}^{\xi},
\label{Abb-rels}
\end{equation}
where either $j=1,\ldots,n_1$, $k=n_1+1,\ldots,n_1+n_2$, $l=1,\ldots,n_1+n_2$ or else
$j=n_1+1,\ldots,n_1+n_2$, $k=1,\ldots,n_1$, $l=1,\ldots,n_1+n_2$.

The parastatistics mixed triple relations for the parafermion and parabosons depend on the family of parafermions under consideration.
For $f_i^\pm$ with $i=1,\ldots,m_1$ and $b_i^\pm$ with $i=1,\ldots,n_1+n_2$, the mixed relations read as follows:
\begin{align}
&[[f_{ j}^{\xi}, f_{ k}^{\eta}], b_{l}^{\epsilon}]=0,\qquad [\{b_{ j}^{\xi}, b_{ k}^{\eta}\}, f_{l}^{\epsilon}]=0, \nn\\
&[[f_{ j}^{\xi}, b_{ k}^{\eta}], f_{l}^{\epsilon}]= -|\epsilon-\xi| \delta_{jl} b_k^{\eta}, \qquad
\{[f_{ j}^{\xi}, b_{ k}^{\eta}], b_{l}^{\epsilon}\}= (\epsilon-\eta) \delta_{kl} f_j^{\xi}.
\label{rel-pf}
\end{align}
On the other hand, for $f_i^\pm$ with $i=m_1+1,\ldots,m_1+m_2$ and $b_i^\pm$ with $i=1,\ldots,n_1+n_2$, the mixed relations read:
\begin{align}
&[[f_{ j}^{\xi}, f_{ k}^{\eta}], b_{l}^{\epsilon}]=0,\qquad
 [\{b_{ j}^{\xi}, b_{ k}^{\eta}\}, f_{l}^{\epsilon}]=0, \nn\\
&\{\{f_{ j}^{\xi}, b_{ k}^{\eta}\}, f_{l}^{\epsilon}\}= 
|\epsilon-\xi| \delta_{jl} b_k^{\eta}, \qquad
[\{f_{ j}^{\xi}, b_{ k}^{\eta}\}, b_{l}^{\epsilon}]= (\epsilon-\eta) \delta_{kl} f_j^{\xi}.
\label{rel-pb}
\end{align}

\end{document}